\documentclass[9pt,conference]{IEEEtran}
\usepackage{dcase2025}

\usepackage{bm} 

\usepackage{dcase2025,amsmath,graphicx,url,times,booktabs, tabularx, multirow, enumitem}
\usepackage{pifont}

\title{On Temporal Guidance and Iterative Refinement \\ in Audio Source Separation}

\name{
Tobias Morocutti$^2$$^*$, Jonathan Greif$^2$$^*$, Paul Primus$^1$$^*$, Florian Schmid$^1$, Gerhard Widmer$^{1,2}$}
\address{$^1$Institute of Computational Perception (CP-JKU),$^2$LIT Artificial Intelligence Lab,\\          
        Johannes Kepler University Linz, Austria \\
        \{tobias.morocutti, jonathan.greif, paul.primus\}@jku.at\\ 
 }

\begin{document}

\maketitle
\def\thefootnote{*}\footnotetext{These authors contributed equally to this work.}\def\thefootnote{\arabic{footnote}}

\begin{abstract}

Spatial semantic segmentation of sound scenes (S5) involves the accurate identification of active sound classes and the precise separation of their sources from complex acoustic mixtures. 
Conventional systems rely on a two-stage pipeline—audio tagging followed by label-conditioned source separation—but are often constrained by the absence of fine-grained temporal information critical for effective separation. 
In this work, we address this limitation by introducing a novel approach for S5 that enhances the synergy between the event detection and source separation stages. 
Our key contributions are threefold. First, we fine-tune a pre-trained Transformer to detect active sound classes. Second, we utilize a separate instance of this fine-tuned Transformer to perform sound event detection (SED), providing the separation module with detailed, time-varying guidance. Third, we implement an iterative refinement mechanism that progressively enhances separation quality by recursively reusing the separator’s output from previous iterations. These advancements lead to significant improvements in both audio tagging and source separation performance, as demonstrated by our system's second-place finish in Task 4 of the DCASE Challenge 2025.
Our implementation and model checkpoints are available in our GitHub repository\footnote{\url{https://github.com/theMoro/dcase25task4}}.

\end{abstract}


\section{Introduction}
\label{sec:intro}
Spatial semantic segmentation of sound scenes (S5) aims to identify and isolate individual sound sources from complex acoustic mixtures.
To achieve this, S5 systems typically operate in two stages. In the first stage, a clip-level event detector identifies the set of sound events present in the mixture. In the second stage, a separator model takes the mixture and one of the predicted class labels as input, and estimates the corresponding isolated source for that class.

For Task 4 of the DCASE 2025 Challenge \cite{yasuda2025}, participants were tasked with training such a system for 18 predefined sound categories, including speech and various domestic sound events. The challenge baseline system~\cite{nguyen2025} employs an M2D-based audio tagger~\cite{M2D} to identify the active sound classes in the mixture (Stage 1), and subsequently separates them using a model based on the ResUNet architecture~\cite{KongResUNet23}, conditioned on the target class via feature-wise linear modulation (FiLM)~\cite{Perez18film} (Stage 2).

\begin{figure*}[t!]
    \centering
    \includegraphics[width=1.0\textwidth]{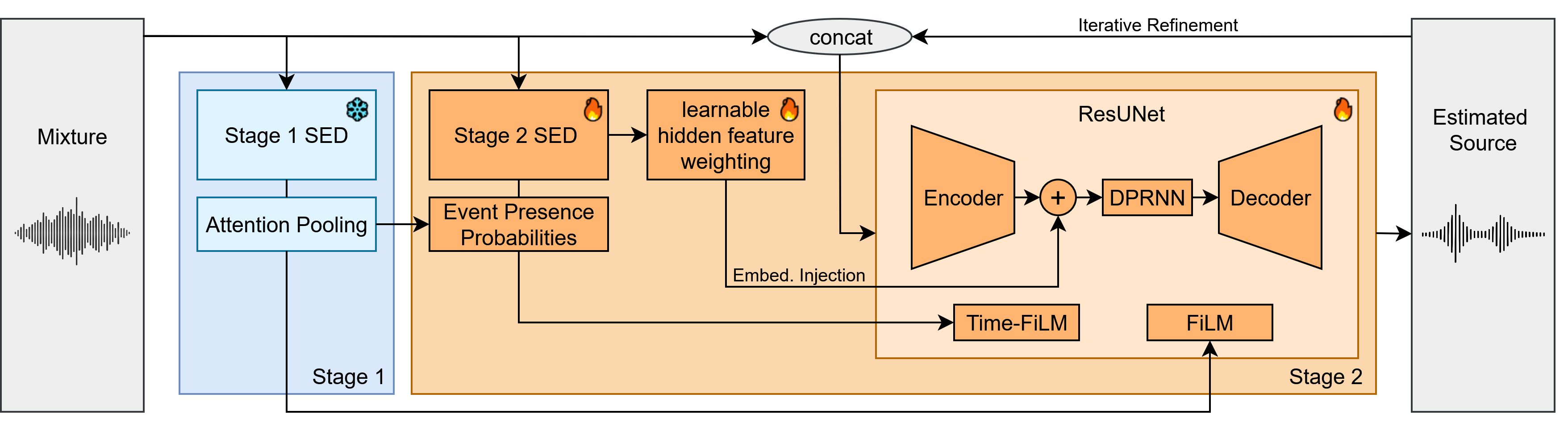}
    \caption{
    Overview of the proposed two-stage audio source separation system. Stage 1 uses a sound event detection (SED) model to identify active sound classes in the mixture. Stage 2 performs class-conditioned source separation using a ResUNet architecture enhanced with temporal conditioning (Time-FiLM), Embedding Injection from the dedicated Stage 2 SED model, and an optional Dual-Path RNN. An iterative refinement mechanism allows the separator to progressively improve output quality by reusing previous predictions as an additional input.
    }
    \label{fig:architecture}
\end{figure*}

In our submission to Task 4 of the DCASE 2025 Challenge \cite{Morocutti_2025_t4}, we enhance the baseline system with several key improvements. This paper presents our system (Figure~\ref{fig:architecture}) and evaluates the impact of our design choices, focusing on the following core areas:

\begin{itemize}
\item \textbf{Sound Event Detection (Stage 1):}
We fine-tune a pre-trained SED Transformer~\cite{Schmid25PretrainedSED}, based on the M2D architecture~\cite{M2D, Niizumi24m2d-clap}, using both clip-level and frame-level annotations for the 18 target classes. At inference time, we use attention-pooled frame-wise predictions to identify active acoustic events in a given mixture.


\item \textbf{Temporal Guidance for Source Separation (Stage 2):}
A trainable copy of the Stage 1 SED model is integrated into the separator to provide temporal guidance. It aids separation by: (1) supplying frame-wise class predictions for enhanced FiLM conditioning (\textit{Time-FiLM}); and (2) injecting a weighted sum of its hidden features into the ResUNet, adding temporally aligned semantics (\textit{Embedding Injection}).

\item \textbf{Dual-Path RNN Integration:}
To better capture long-range temporal dependencies, we augment the ResUNet’s embedding space with a Dual-Path RNN~\cite{Luo20dprnn}, enabling more effective context modeling across time.

\item \textbf{Iterative Refinement:}
Inspired by recent iterative separation strategies \cite{Yuan_FlowSep}, we apply a simple yet effective refinement technique: the separator’s outputs are recursively fed back into the model, progressively improving the separation quality.
\end{itemize}

Section~\ref{sec:formalization} provides a brief formalization of the problem. In Sections~\ref{sec:class_detection} and~\ref{sec:sep}, we describe our modifications to the event detection and separation stages, respectively. Section~\ref{sec:exp_setup} outlines the experimental setup, and Section~\ref{sec:results} presents the results of our investigation into the effectiveness of the proposed enhancements. Finally, Section~\ref{sec:conclusion} concludes this paper. 

\section{Problem Formalization}
\label{sec:formalization}

The goal of the \textit{S5} task is to detect and separate individual sound events from multi-channel time-domain mixtures recorded in realistic environments. Let
\[
\boldsymbol{X} = [\boldsymbol{x}^{(1)}, \dots, \boldsymbol{x}^{(M)}]^\top \in \mathbb{R}^{M \times T}
\]
be the multi-channel mixture of length $T$, recorded with $M$ microphones. Each channel $\boldsymbol{x}^{(m)} \in \mathbb{R}^T$ is modeled as:

\begin{equation}
\label{eq:task_setting}
\begin{split}
\boldsymbol{x}^{(m)}  & = \sum_{k=1}^{K} \boldsymbol{h}_k^{(m)} * \boldsymbol{s}_k +  \sum_{j=1}^{J} \boldsymbol{h}_j^{(m)} * \boldsymbol{s}_j + \boldsymbol{n}^{(m)} \\
\end{split}
\end{equation}

where $*$ denotes the convolution operator, $K$ is the number of active target sound sources, and $J$ is the number of interference events.
The terms $\boldsymbol{s}_k, \boldsymbol{s}_j \in \mathbb{R}^T$ represent the single-channel dry source signals for a target event of class $c_k$ and an interference event of class $c_j$, respectively. 
Similarly, $\boldsymbol{h}_k^{(m)}, \boldsymbol{h}_j^{(m)} \in \mathbb{R}^H$ are the room impulse responses (RIRs) from the target source $k$ and interfering source $j$ to microphone $m$, respectively, where $H$ denotes the RIR length.
Finally, $\boldsymbol{n}^{(m)} \in \mathbb{R}^T$ represents additive noise on the $m$-th microphone channel.

The objective is to estimate the active target classes $c_k$ and recover their direct-path waveforms $\{\boldsymbol{s}^{(d,m)}_1, \dots, \boldsymbol{s}^{(d,m)}_K\}$ for a chosen reference microphone channel $m$, where

\begin{equation}
\boldsymbol{s}^{(d, m)}_k = \boldsymbol{h}_{{k}}^{(d,m)} * \boldsymbol{s}_k
\end{equation}

is the anechoic source $\boldsymbol{s}_k$ convolved with the direct-path RIR $\boldsymbol{h}_k^{(d,m)}$, corresponding to the direct arrival sound without reverberation. 

For our purposes, $\boldsymbol{h}_k^{(d,m)}$ is approximated from the full RIR $\boldsymbol{h}_k^{(m)}$ by applying a time-domain window around the first significant energy peak. 


\section{Active Event Detection}
\label{sec:class_detection}

Stage~1 relies on an audio classifier to estimate the presence or absence of the $C$ target acoustic events at the clip level. Unlike the baseline system, which employs a standard audio tagger, we use a sound event detection (SED) model that predicts class activity at the frame level. These predictions are then aggregated into clip-level class scores via attention pooling. This design choice is motivated by two key factors: (1) we hypothesize that frame-level supervision offers more informative training signals, potentially improving tagging accuracy, and (2) the resulting SED model can be integrated into the separator in Stage~2 to provide temporal conditioning.

For training the SED model, we use only the first channel of the multi-channel mixture, denoted $\boldsymbol{x}^{(1)}$, and convert it into a mel-spectrogram:
\[
\{\mathbf{y}_t\}_{t=1}^{T} = \mathrm{MEL}(\boldsymbol{x}^{(1)}), \quad \mathbf{y}_t \in \mathbb{R}^F,\
\]
where $T$ is the number of time frames and $F$ is the number of mel frequency bins. The model, denoted by a function $g$, processes this input and produces a sequence of embedding vectors:
\begin{equation}
\label{eq:forward}
\{\hat{\mathbf{e}}_s\}_{s=1}^{S} = g\big(\{\mathbf{y}_t\}_{t=1}^{T}\big), \quad \hat{\mathbf{e}}_s \in \mathbb{R}^D,
\end{equation}
where $D$ is the embedding dimension, and $S$ is the output sequence length. Note that $S$ may differ from $T$ due to subsampling within the model architecture.

Next, frame-level (strong) class prediction probabilities are computed using a linear layer followed by a sigmoid activation:
\begin{equation}
\label{eq:strong}
\{\hat{\mathbf{o}}^{(\mathrm{strong})}_s\}_{s=1}^{S}\ = \sigma\big( \mathbf{W_h} \{\hat{\mathbf{e}} \}_{s=1}^{S} + \mathbf{b_h} \big),
\end{equation}
where $\mathbf{W_h} \in \mathbb{R}^{C \times D}$ and $\mathbf{b_h} \in \mathbb{R}^C$ are the learnable weight matrix and bias vector of the output layer, and $\sigma$ is the element-wise sigmoid function. To obtain clip-level (weak) predictions $\hat{\mathbf{o}}^{(\mathrm{weak})}$ from the frame-level outputs $\{\hat{\mathbf{o}}^{(\mathrm{strong})}_s\}_{s=1}^{S}$, we compute class-specific attention weights over time. These are generated using a separate attention head:
\begin{equation}
\label{eq:attention}
\{\boldsymbol{\alpha}_s\}_{s=1}^{S} = \textrm{softmax}\big( \mathbf{W_a} \{ \hat{\mathbf{e}}_s\}_{s=1}^S + \mathbf{b_a} \big),
\end{equation}
where $\mathbf{W_a} \in \mathbb{R}^{C \times D}$ and $\mathbf{b_a} \in \mathbb{R}^C$ are the weight matrix and the bias vector of the attention head, and the softmax function is applied independently for each class across time steps. The final clip-level prediction is then obtained as the attention-weighted average of the strong outputs:
\[
\hat{\mathbf{o}}^{(\mathrm{weak})} = \sum_{s=1}^{S} \boldsymbol{\alpha}_s \cdot \hat{\mathbf{o}}^{(\mathrm{strong})}_s.
\]

We train the model using a loss function that combines binary cross-entropy (BCE) losses for both strong and weak labels, balanced by a weighting factor $\lambda$:
\begin{equation}
\label{eq:loss}
\begin{split}
\mathcal{L} =\ & \lambda \cdot \frac{1}{SC} \sum_{s=1}^{S} \sum_{c=1}^{C} 
\mathrm{BCE}\big( \hat{o}^{(\mathrm{strong})}_{s,c},\ y^{(\mathrm{strong})}_{s,c} \big) \\
&+ (1 - \lambda) \cdot \frac{1}{C} \sum_{c=1}^{C} 
\mathrm{BCE}\big( \hat{o}^{(\mathrm{weak})}_{c},\ y^{(\mathrm{weak})}_{c} \big).
\end{split}
\end{equation}

In all experiments, we set $\lambda = 0.5$.


\section{Source Separation}
\label{sec:sep}


The second stage employs a modified ResUNet architecture~\cite{KongResUNet23, nguyen2025}, which takes the multi-channel mixture waveform and a target class label as input to produce an estimated single-channel source.
The ResUNet operates on magnitude spectrograms and predicts magnitude and phase masks to reconstruct the target’s complex spectrogram, which is then inverted to the time domain via the inverse Short-Time Fourier Transform.

\subsection{Time-FiLM Conditioning}

In addition to standard clip-level conditioning via Feature-wise Linear Modulation (FiLM)~\cite{Perez18film}, we provide finer-grained temporal guidance by integrating a dedicated and trainable SED model---referred to as the Stage~2 SED model---into the separator architecture. We refer to this approach as \textit{Time-FiLM} as it generalizes FiLM~\cite{Perez18film} by incorporating a temporal dimension. The Stage~2 SED model is initialized with the weights of the fine-tuned model described in Section~\ref{sec:class_detection}, but is subsequently optimized jointly with the separator.

For a class predicted by the Stage~1 SED model, the corresponding class activity probability map obtained by the Stage~2 SED model is passed through a feedforward network to generate a sequence of embedding vectors. These embeddings are transformed into a sequence of channel-wise scale and shift parameters for each conditioned layer in the ResUNet. To align those sequences with the feature maps, the sequences are interpolated in time. Finally, Time-FiLM modulation is applied by scaling and shifting the feature maps with their time-aligned scale and shift parameters.


\subsection{Embedding Injection}

To further enhance separation quality, we inject intermediate representations from the Stage~2 SED model into the ResUNet’s latent space. Inspired by~\cite{olivan25soundbeam}, we extract hidden representations after each of the \(N\) blocks in the Stage~2 SED model, denoted by \(Z_{\text{TF}}^{(i)} \in \mathbb{R}^{C_{\text{TF}} \times F_{\text{TF}}' \times T_{\text{TF}}'}\), and compute a weighted combination using learnable weights:

\begin{equation}
\label{eq:weighted_sum}
\alpha_i = \frac{\exp(w_i)}{\sum_{j=1}^{N} \exp(w_j)}, \quad
Z_{\text{TF}} = \sum_{i=1}^{N} \alpha_i \cdot Z_{\text{TF}}^{(i)},
\end{equation}

where \(w_i\) are learned scalar parameters. The resulting fused representation \(Z_{\text{TF}}\) is projected through a linear transformation and then interpolated along both the frequency and time dimensions to match the shape of the ResUNet’s latent features \(Z_{\text{RN}} \in \mathbb{R}^{C_{\text{RN}} \times F_{\text{RN}}' \times T_{\text{RN}}'}\). 
After normalizing both the fused representation and the ResUNet’s latent features, we add them element-wise. 

\subsection{Dual-Path RNN}

To capture long-range dependencies, we integrate a Dual-Path Recurrent Neural Network (DPRNN)~\cite{Luo20dprnn} into the ResUNet’s embedding space. The DPRNN consists of two blocks, each with two stacked bidirectional GRUs~\cite{GRU}: 
the first processes along the temporal dimension, the second along the frequency dimension.

\subsection{Iterative Refinement}

Instead of generating the target source in a single forward pass, we treat separation as an iterative process. At each step, the estimated single-channel output is stacked with the original \(M\)-channel mixture to form a new \((M+1)\)-channel input. The ResUNet’s first convolution is adapted for \( M+1 \) channels, with the extra channel initialized to zeros during the first iteration.

This iterative mechanism allows the separator to refine its predictions over multiple steps. During training, the number of iterations is randomly sampled between 1 and a predefined maximum. To reduce the memory footprint during training, gradients are detached between iterations and only the final step contributes to parameter updates.


\section{Experimental Setup}
\label{sec:exp_setup}

This section details the experimental setup, including the dataset, evaluation metric, model configurations, and the design used to assess our key contributions.

\subsection{Dataset}
\label{subsec:dataset}

Training and validation use the DCASE 2025 Task 4 (S5) development set, with 4-channel 10-second mixtures synthesized on-the-fly at 32kHz using a modified SpatialScaper library~\cite{romain24spatial_scaper}. Training data is randomly sampled, while validation mixtures are generated deterministically. Each mixture combines the following four components:

\begin{itemize}
    \item \textbf{Target sound events} $\boldsymbol{s}_k$:
    Anechoic recordings from 18 target classes (e.g., \textit{AlarmClock}, \textit{FootStep}, \textit{Speech}) are used, mainly from FSD50K~\cite{FSD50K} and EARS~\cite{Riachter24EARS}, with additional data from NTT. The training set includes 7,336 samples (33–2,063 per class), and the validation set has 317 samples (1–37 per class), showing significant class imbalance.
    \item \textbf{Interference events} $\boldsymbol{s}_j$:
    We utilize isolated recordings of 94 non-target classes from the Semantic Hearing dataset~\cite{Veluri23semantic_hearing}. This set includes 6,184 training and 845 validation samples, displaying class distribution inbalance similar to the target sound events.
    \item \textbf{Room impulse responses} $\boldsymbol{h}$: 
    For acoustic environment simulations we employ multichannel RIRs captured from various real indoor locations. Our setup incorporates 5 RIRs for training and 3 for validation. The recordings are partially sourced from the FOA-MEIR dataset~\cite{yasuda22foa-meir}, with additional samples newly recorded by NTT.
    \item \textbf{Ambient noise} $\boldsymbol{n}$: We include real-world background noise recordings obtained from the FOA-MEIR dataset~\cite{yasuda22foa-meir}. The noise dataset contains 59 training and 21 validation samples. 
\end{itemize}

This synthetic setup also provides precise event onsets and offsets, which are required for training the SED models (see Section~\ref{sec:class_detection}).  For testing, we use the pre-synthesized mixtures provided by the DCASE challenge organizers.
For all sets, each mixture contains between one and three target events and up to two interfering events. The target events have per-event Signal-to-Noise Ratios (SNRs) ranging from 5 to 10 dB, while interfering events have SNRs between 0 and 15 dB. 

\subsection{Evaluation Metric}

We evaluate our system using Class-Aware Signal-to-Distortion Ratio improvement (CA-SDRi)~\cite{nguyen2025}, a metric that jointly measures source separation quality and classification accuracy. Let $C$ be the set of ground truth classes and $\hat{C}$ the set of predicted classes. For correctly predicted classes $c_k \in C \cap \hat{C}$, we compute the SDR improvement between the estimated signal $\hat{\boldsymbol{s}}_k$ and the corresponding target source relative to the mixture's first channel $x^{(1)}$. For simplicity, we denote the target direct-path signal $\boldsymbol{s}_k^{(d,m)}$ as $\boldsymbol{s}_k$ in the following equations.

\begin{equation}
\text{SDRi}(\hat{\boldsymbol{s}}_k, \boldsymbol{s}_k, \boldsymbol{x}^{(1)}) = \text{SDR}(\hat{\boldsymbol{s}}_k, \boldsymbol{s}_k) - \text{SDR}(\boldsymbol{x}^{(1)}, \boldsymbol{s}_k)
\end{equation}

with SDR defined as:

\begin{equation}
\text{SDR}(\hat{\boldsymbol{x}}, \boldsymbol{x}) = 10 \log_{10} \left( \frac{|\boldsymbol{x}|^2}{|\boldsymbol{x} - \hat{\boldsymbol{x}}|^2} \right)
\end{equation}

The final CA-SDRi score is computed by averaging over all unique classes in $C \cup \hat{C}$:

\begin{equation}
\text{CA-SDRi} = \frac{1}{|C \cup \hat{C}|} \sum_{c_k \in C \cup \hat{C}} P_{c_k},
\end{equation}

where $P_{c_k} = \text{SDRi}(\hat{\boldsymbol{s}}_k, \boldsymbol{s}_k, \boldsymbol{x}^{(1)})$ if $c_k \in C \cap \hat{C}$, and $P_{c_k} = 0$ otherwise.

\subsection{Class Detection Model}
\label{subsec:exp_sed}

The model used for detecting active acoustic events is based on the M2D architecture~\cite{M2D, Niizumi24m2d-clap}, pre-trained for audio tagging on AudioSet~\cite{Gemmeke17AudioSet}. We use a checkpoint further fine-tuned for sound event detection (SED) on temporally strong labels from AudioSet Strong~\cite{Hershey21ASstrong}, following the procedure in~\cite{Schmid25PretrainedSED}. This model is chosen for its state-of-the-art SED performance, as shown in~\cite{Schmid25PretrainedSED}.
The model is trained as outlined in Section~\ref{sec:class_detection} for 22,000 steps with a batch size of 32. We employ a cosine learning rate schedule with 4,000 warm-up steps and utilize the ADAM optimizer~\cite{ADAM_Kingma} with no weight decay. The initial learning rate is set to $1\!\times\!10^{-3}$, and a layer-wise learning rate decay factor of 0.8 is applied to adjust the learning rates across different layers of the model.

\subsection{Source Separation Model}
\label{subsec:exp_sep}


We use a ResUNet architecture~\cite{Resunet, KongResUNet23} for sound source separation, operating at 32kHz with a window size of 2048 and a hop size of 160 samples, matching the baseline. The model is initialized with pre-trained weights from AudioSep~\cite{Liu23AudioSep}, a language-conditioned separation framework. While AudioSep was originally trained with a 320-sample hop size, we found a hop size of 160 samples to perform better in our setup.
For fine-tuning the separator, we use separate learning rates: $2\!\times\!10^{-4}$ for pre-trained components, $5\!\times\!10^{-4}$ for DPRNN parts and $1\!\times\!10^{-3}$ for newly introduced 
parts such as the Time-FiLM embedding layers. Additionally, we train the Stage~2 SED model using a learning rate of $4 \! \times \! 10^{-4}$ and a learning rate decay factor of 0.9. 
Training runs for 450,000 steps with an batch size of 8, using the Adam optimizer~\cite{ADAM_Kingma} (no weight decay) and a cosine learning rate schedule with 12,000 warm-up steps.


\subsection{Experimental Design}
\label{subsec:exp_design}

To assess the efficacy of the proposed techniques for enhancing source separation, we conduct a series of experiments structured as follows. Initially, we evaluate the impact of Time-FiLM and Embedding Injection, both of which utilize the Stage~2 SED model. We also investigate the benefits of allowing this model to be trainable during the training phase of the separation model. We refer to the configuration incorporating Time-FiLM, Embedding Injection, and a trainable Stage~2 SED model as \textit{AudioSep-SED}. This model serves as the foundation for subsequent experiments. Building upon it, we integrate a Dual-Path RNN (DPRNN) into the latent space to potentially enhance the model's ability to capture long-range dependencies.
Subsequently, we explore the iterative refinement approach by training models with varying numbers of iterations (2, 3, and 4) and assessing their performance when employing up to 10 iterations during inference.

\section{Results}
\label{sec:results}


The Stage 1 sound event detection model proposed in Section~\ref{sec:class_detection} increases test set accuracy by 8 points over the DCASE baseline, from 59.8\% to 67.8\%.

\begin{table}[t]
\small 
\setlength{\tabcolsep}{4pt} 
\begin{center}
\begin{tabular}{l@{\hskip 6pt}c@{\hskip 6pt}c@{\hskip 6pt}c}
\toprule
\textbf{Config} & \textbf{M2D \cite{nguyen2025}} & \textbf{M2D (Ours)} & \textbf{Oracle} \\
\midrule
DCASE Baseline~\cite{nguyen2025} & 11.03 & -- & -- \\
\midrule
AudioSep-SED & \textbf{12.71$\pm$0.03} & \textbf{13.42$\pm$0.11} & \textbf{15.29$\pm$0.13} \\
- w/o Time-FiLM & 12.41$\pm$0.03 & 13.07$\pm$0.06 & 14.82$\pm$0.06 \\
- w/o Embedding Injection & 12.46$\pm$0.01 & 13.12$\pm$0.07 & 14.91$\pm$0.09 \\
- w/o Trainable S2-SED & 11.95$\pm$0.04 & 12.53$\pm$0.03 & 14.04$\pm$0.04 \\
\midrule
AudioSep-SED + DPRNN & 12.55$\pm$0.07 & 13.31$\pm$0.07 & 15.20$\pm$0.12 \\
\bottomrule
\end{tabular}
\vspace{6pt}
\caption{
CA-SDRi scores (↑) for different system variants on the development test split, where \textit{AudioSep-SED w/o Trainable S2-SED} represents the AudioSep-SED configuration, where the Stage~2 SED model is not trainable. Columns correspond to class detection using the baseline M2D, our fine-tuned M2D, and oracle targets. Bold indicates the best per column. 
}
\label{tab:ablation}
\end{center}
\end{table}



Table~\ref{tab:ablation} compares the performance of the DCASE baseline, AudioSep-SED, and AudioSep-SED + DPRNN, highlighting the impact of each technique on the test split of the DCASE 2025 S5 development set. Performance is measured using CA-SDRi under three Stage 1 models: (1) baseline M2D tagger, (2) our M2D model, and (3) oracle targets. Results are averaged over two runs.

The second section in Table~\ref{tab:ablation} shows that AudioSep-SED outperforms the DCASE baseline system. Integrating Time-FiLM, Embedding Injection, and a trainable Stage~2 SED model improves performance regardless of Stage~1 predictions, with the trainable Stage~2 SED yielding the largest gain. While Time-FiLM contributes more than Embedding Injection, their combination with a trainable SED achieves the best results.
With baseline M2D predictions, this setup reaches a CA-SDRi of 12.71 vs. 11.03 for the DCASE baseline. Using our M2D model, the score further increases to 13.42.



\begin{figure}[t!]
    \includegraphics[width=\columnwidth]{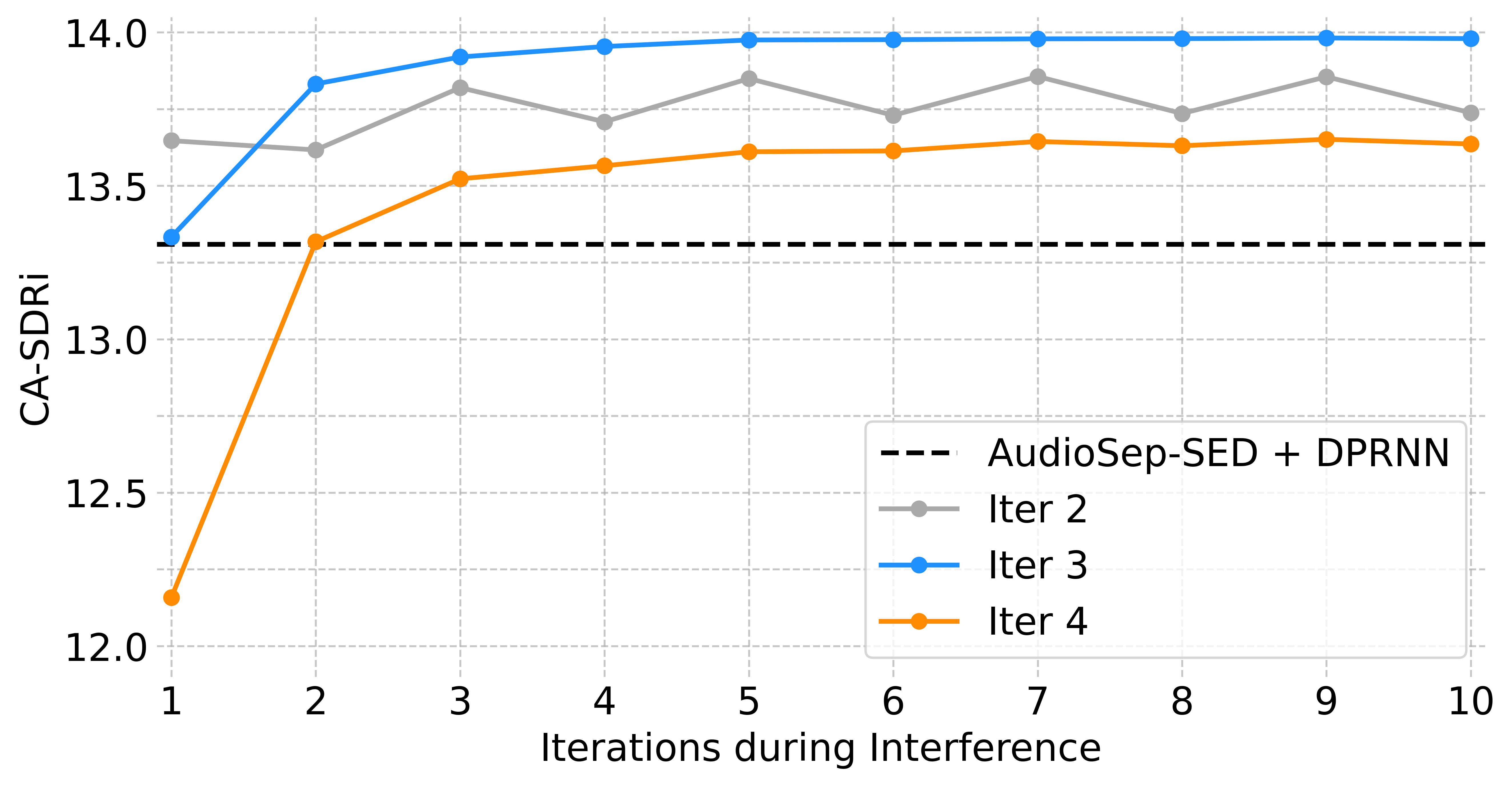}
    \caption{
Performance of the iterative refinement experiments compared to the AudioSep-SED + DPRNN configuration evaluated on the test set.
\textit{Iter \{2,3,4\}} denotes experiments with the maximum training iterations set to 2, 3, or 4, respectively. Class detection is performed using our fine-tuned M2D model.
    }
    \label{fig:plot2}
    \vspace{-4pt}
\end{figure}

Adding a DPRNN to AudioSep-SED improves performance on the \textit{validation split}, increasing CA-SDRi from 13.79 to 14.35 using baseline M2D predictions. However, it underperforms on the \textit{test split} (third section in Table~\ref{tab:ablation}), suggesting a distribution mismatch. Following standard practice, we select the best validation setup for further experiments; thus, DPRNN is included in our iterative refinement runs.


Figure~\ref{fig:plot2} shows the performance of models trained with up to 2, 3, or 4 iterations and evaluated using 1–10 iterations at inference. Iterative refinement clearly outperforms the AudioSep-SED + DPRNN setup, with most gains occurring in early iterations and diminishing improvements thereafter—indicating convergence after a few cycles. Notably, using more inference iterations than used during training can still boost performance; e.g., a model trained with 3 iterations reached 13.92 after 3 inference steps and 13.98 after 10.


Our iterative refinement models generally enhance performance as the number of inference iterations increases, although the performance changes between consecutive iterations can be inconsistent and occasionally negative. A notable pattern emerges where odd-numbered iterations frequently outperform their even-numbered counterparts, a trend most pronounced in models trained with two iterations (Iter 2). However, this odd-even performance gap narrows in models trained with three or four iterations (Iter 3 and Iter 4), which suggests that extending the training iterations helps stabilize the refinement process.

Furthermore, Iter 3 outperforms both Iter 2 and Iter 4, indicating it strikes a good balance between training complexity and output quality. 
That said, because these experiments relied on a single run per configuration, the observed results should be interpreted with caution, as they may be influenced by random variability.




\section{Conclusion}
\label{sec:conclusion}
We introduced a two-stage audio source separation system that leverages sound event detection (SED) models to (1) identify active events and (2) guide the separator via temporal conditioning (Time-FiLM) and Embedding Injection. Evaluated on the DCASE 2025 S5 dataset, our results show that SED improves event detection in Stage~1 and provides effective guidance for Stage~2 separation. We also proposed an iterative refinement strategy, where outputs are recursively fed back into the model, which progressively enhanced the separation quality.

\section{ACKNOWLEDGMENT}
\label{sec:ack}


The computational results presented were obtained in part using the Austrian Scientific Cluster (ASC) and the Linz Institute of Technology (LIT) AI Lab Cluster. The LIT AI Lab is supported by the Federal State of Upper Austria. Gerhard Widmer's work is supported by the European Research Council (ERC) under the European Union's Horizon 2020 research and innovation programme, grant agreement No 101019375 (Whither Music?).



\bibliographystyle{IEEEtran}
\bibliography{refs}







\end{document}